\documentclass[12pt]{article}

\begin{document}

\title{Solvent mode participation in the non-radiative relaxation 
of the hydrated electron}

\author{Oleg V. Prezhdo and Peter J. Rossky\\ \\ 
\em Department of Chemistry and Biochemistry\\
\em University of Texas at Austin\\
\em Austin, Texas 78712-1167}

\date{Submitted to J. Phys. Chem.: \today}

\maketitle


\begin{abstract}
\setlength{\baselineskip}{7mm}    
Non-adiabatic molecular dynamics simulations are used
to analyze the role of different solvent degrees of freedom in
the non-radiative relaxation of the first excited state of the 
hydrated electron. The relaxation occurs through a multi-mode
coupling between the adiabatic electronic states.  The process 
cannot be described by a single mode promotion model frequently
used in the ``large molecule'' limit of gas phase theories.
Solvent librations and vibrations, and the H$_2$O asymmetric stretch
in particular, are found to be the most effective promotors of 
the electronic transition. Dissipation of the released energy to 
the solvent proceeds on two time scales: a fast 10-20~fs heating of the
first solvation shell, where most of the energy is accepted by the librational
degrees of freedom, and a several hundred femtosecond global reconstruction of the
solvent as the first shell transfers its excess energy to the rest
of the molecules.  The implications of our use of a semiclassical approximation
as the criterion for good promoting and accepting modes are discussed.
\end{abstract}


\setlength{\baselineskip}{7mm}    

\section{Introduction}\label{sec1}

The majority of chemical processes involve radiationless energy transfer
in one or more of their elementary steps, be it an intermolecular conversion,
intramolecular relaxation, or an ordinary heating necessary to activate a
reaction. The role of non-radiative processes is particularly pronounced
in liquid phase chemistry, where the solvent serves as a thermal bath
providing or withdrawing energy from the reacting species.  During the last
decade new experimental and theoretical techniques have been developed allowing
for detailed dynamical studies on ultrafast processes in polar solutions.
The hydrated electron, being a relatively simple as well as ubiquitous
species in solution photo- and electrochemistry, represents a unique system for
such studies.  Because the hydrated electron is merely an extra electron
in pure water, its evolution is governed solely by nuclear motions of
water molecules.  The instantaneous configuration of the solvent determines
the Born-Oppenheimer (B-O) electronic energy spectrum. Vibrations and librations 
couple these electronic states leading to non-adiabatic radiationless
transitions.  The electronic energy released during the transitions is accepted 
by specific solvent modes, and, as the solvation structure
approaches equilibrium, the energy is dissipated into the rest of the solvent.
The present work was carried out to elucidate in detail 
what particular solvent motions and solvent regions have 
the greatest impact on these aspects of hydrated electron relaxation dynamics. 
This analysis represents the first such detailed examination for an electronic
relaxation in solution.

The idea of non-radiative paths of energy transfer between
states of two species was born in photon absorption and emission 
experiments.  The first recorded
evidence of radiationless processes dates back to 1923, when Cario and 
Frank$^1$ studied emission spectra of mercury--thallium gaseous
mixtures irradiated by light resonant with an electronic transition
of mercury atoms.  Emission spectra showed both mercury and thallium
lines suggesting that a fraction of mercury atoms non-radiatively
transferred their excess energy to thallium atoms. 
Experimental studies on radiationless relaxation in isolated molecules were
first carried out in the 1960's.$^{2,3}$  The results
were interpreted using a molecular energy level scheme based on the Born-Oppenheimer
separability of electronic and nuclear degrees of freedom. Non-radiative
relaxation of excited adiabatic electronic states was explained in terms
of vibronic coupling treated perturbatively (Fermi Golden 
Rule).$^{4-7}$ The complexity of the problem lies 
in the huge dimensionality of vibronic basis sets, and often further approximations are
involved, the most well-known being the introduction of Franck-Condon factors. 
Intramolecular radiationless processes within relatively small isolated molecules 
were extensively studied both experimentally and 
theoretically.$^{8-12}$
Several new concepts were introduced, such as promoting modes -- those which
induce non-radiative electronic transitions between B-O states,
and accepting modes -- those that accommodate 
the energy released during such transitions, when these occur at
points away from curve crossings.  Depending on the number of
vibrations that couple electronic states, ``large'' and ``small'' molecule
limits were suggested.  Many theoretical techniques were used to treat vibronic 
Hamiltonians: Laplace transformation, Green's function, projection
operator, effective Hamiltonian methods, and wave packet formalism.$^8$

Molecular energy transfer in condensed media is a less developed 
subject$^{13,14}$ and is notably based 
on the gas phase theories.  Experiments on luminescence of large molecules in solutions
were conducted$^{15-17}$ prior to the corresponding 
experiments in gas phase.$^{2,3}$ However, they
provided less detailed data. Theoretical
description of the results of condensed phase measurements is complex due to
the presence of a solvent which both accepts released energy and contributes 
to relaxation promotion.  In the case
of monoatomic solutes (including the hydrated electron) solvent motion alone
is responsible for the coupling between adiabatic electronic states.
In the simplest form the solvent can be described
by its macroscopic characteristics (viscosity and dielectric response)
leading to continuum models. Although such models still play a major
role in some areas$^{18-20}$, 
they cannot reproduce ultrafast solvent dynamics
on the time scale of motions of individual molecules. Molecular dynamics
methods consider solute-solvent systems in their full molecularity
and thus are more suitable for simulations of energy transfer processes
in solutions. A short time approximation to the evolution of molecular
trajectories leads to another ansatz -- instantaneous normal 
modes.$^{21,22}$  At each instance of time, the solvent is regarded as 
a single large molecule characterized by a quasi-continuous spectral density of
vibrational states.  The model of a harmonic bath coupled to a solute
was successful in a generic description of vibrational relaxation in condensed 
media$^{13}$, and harmonic baths are also widely used
in other areas, such as quantum decoherence$^{23,24}$, 
tunneling$^{25,26}$ and Brownian motion.$^{27}$

Recent advances in time resolved spectroscopies$^{28-33}$, and parallel theoretical 
research$^{34-42}$ has made the aqueous solvated electron an exceptional probe 
of dynamics of chemical processes in polar solvents.$^{43-45}$
The hydrated electron is a relatively simple system from the theoretical point
of view.  It is also tractable
by spectroscopic techniques due to its large optical cross-section.
At the same time, the hydrated electron exhibits a remarkably rich
dynamical picture: the hydrated electron excited electronic state relaxation
is intimately intertwined with the configurational rearrangement of the solvent
as evidenced by recent experimental studies$^{32,33}$
involving photoexcitation of the equilibrium ground state and subsequent 
nonradiative relaxation back to the ground state, as well as by
earlier experiments$^{28-31}$ on hydrated electrons created by a multiphoton 
ionization of pure water at much higher energies.
Many aspects of the electron evolution have been elucidated via theoretical
analysis employing simulated dynamics and comparison to 
experiment.$^{46-53}$

The present paper focuses on the previously unexplored aspect associated with
the identification of particular modes of water molecules
that most strongly couple adiabatic states of the hydrated electron
(promoting modes), and those that provide effective sinks for the
energy released during non-adiabatic radiationless transitions (accepting modes), 
and also during the subsequent adiabatic relaxation of the electron.  The next
section deals with the criteria for good promoting and accepting modes and the
implications of our use of a semiclassical approach.  We then discuss in 
Sec.~\ref{sec3} the excess-electron-in-water system in detail 
and, finally, summarize our conclusions in Sec.~\ref{sec4}.


\section{Theory}\label{sec2}

The results here are obtained from simulations based on 
a variant of non-adiabatic molecular dynamics
devised by Webster {\em et al.}$^{37}$, which
incorporates a semiclassical force on the solvent$^{54}$
derived via a path integral formulation$^{55}$
and employs a surface hopping technique~$^{56,57}$ for determining 
transition rates.
For our studies we use the results of earlier simulations.$^{48}$
The model consists of 200 classical SPC flexible water 
molecules$^{58,59}$
and an electron treated quantum mechanically. The electron--water 
pseudopotential was taken from earlier work.$^{60}$
Twenty non-adiabatic trajectories were considered, characterized 
by solvent density 0.997~g/cm$^3$ at room temperature.  
Each of the trajectories was initiated from a configuration of an adiabatically
equilibrated electron in its ground state. The electron was then
instantaneously promoted to the first excited state 
and allowed to non-adiabatically relax back to the ground state. 


\subsection{Separation of solvent contributions by modes and space}\label{sec2aa}

To draw physically clear conclusions about the role of the solvent
in the electron relaxation, a convenient coordinate system for the solvent degrees
of freedom is needed.  
We use the following scheme to separate the overall solvent motion into the subsets 
of internal motions of individual molecules.
For each water molecule we define the translational degrees of
freedom by separating the center of mass motion and the three rotational modes 
by diagonalizing the instantaneous moment of inertia tensor.
To define the vibrational modes one could
expand the potential acting on the molecule up to second order terms
in the displacement from the instantaneous configuration.
Diagonalization of the corresponding Hessian matrix would produce the normal modes 
and the vibrational frequencies.  However, a water molecule in solution 
can be substantially distorted (in our simulations the 
H--O--H angle and O--H bond length varied from 90 to 120 degrees and from
0.95 to 1.15 \AA~respectively) and is also subject to a strong external field.
In general, this normal mode procedure would lead to different sets of 
modes for different molecules with no trivial correspondence between
the sets. For these reasons, we use an approximate scheme to factor out distinct
vibrational modes in order to obtain equivalents of the symmetric and 
asymmetric stretching and bending motions of an undistorted water molecule.
The scheme was motivated by the normal mode analysis of 
a non-linear ABA molecule carried out in reference~\cite{Landafshits.Mechanics}.

The three dimensional vibrational subspace can be written in terms of projector
operators as $P_{vib}=I-P_{tr}-P_{rot}$.
Transforming the Cartesian coordinates of the water molecule to
the frame where the moment of inertia tensor is diagonal and the
molecule is in the x-z plane (Fig.~\ref{fig:vib}) we split the
projector on the vibrational subspace into the contributions
from the three orthonormal modes $P_{vib}=IP_{vib}=(P_{as}+P_{ss}+P_{b})P_{vib}$
roughly corresponding to the asymmetric stretch ($P_{as}$), symmetric stretch
($P_{ss}$) and bend ($P_{b}$)
\begin{eqnarray}
P_{as}&=&|X_{O} \rangle \langle X_{O}| + |X_{H}^{a} \rangle \langle X_{H}^{a}| 
\nonumber \\
P_{ss}&=&|c_{+} \rangle \langle c_{+}| + |d_{+} \rangle \langle d_{+}| \nonumber \\
P_{b}&=&|c_{-} \rangle \langle c_{-}| + |d_{-} \rangle \langle d_{-}| 
\label{eq:vib}
\end{eqnarray}
where the auxiliary coordinates 
\begin{eqnarray}
|X_{H}^{s} \rangle &=&(|X_{H_{1}} \rangle -|X_{H_{2}} \rangle ) /\sqrt{2} \nonumber \\
|X_{H}^{a} \rangle &=&(|X_{H_{1}} \rangle +|X_{H_{2}} \rangle ) /\sqrt{2} \nonumber \\
|Z_{H}^{s} \rangle &=&(|Z_{H_{1}} \rangle -|Z_{H_{2}} \rangle ) /\sqrt{2} \nonumber \\
|Z_{H}^{a} \rangle &=&(|Z_{H_{1}} \rangle +|Z_{H_{2}} \rangle ) /\sqrt{2} \nonumber 
\end{eqnarray}
are the symmetrized and antisymmetrized Cartesian displacements of the
hydrogens, and
\begin{eqnarray}
|c_{\pm} \rangle &=&(|Z_{H}^{a} \rangle  \pm |X_{H}^{s} \rangle ) /\sqrt{2} \nonumber \\
|d_{\pm} \rangle &=&(|Z_{O} \rangle  \pm |Z_{H}^{s} \rangle ) /\sqrt{2} \nonumber 
\end{eqnarray}
Here, $X_{i}$ and $Z_{i}$, $i=O, H_{1}, H_{2}$ denote the Cartesian displacements 
of the {\em i}-th atom along the $x$ and $z$-axes (Fig.~\ref{fig:vib}). 
The asymmetric stretch projector picks out all ``antisymmetric'' movements
along the $x$-axis.  The sum of the symmetrized displacements 
$|Z_{H}^{a} \rangle +|X_{H}^{s} \rangle $ forms the base of the symmetric stretch, while
their difference $|Z_{H}^{a} \rangle -|X_{H}^{s} \rangle $ contributes to the bend.
Figure~\ref{fig:vib} shows the vibrations defined above.
The projectors (Eq.~\ref{eq:vib}) have the necessary properties 
to make the vibrational modes defined by these projectors form 
an orthonormal basis, namely,
$P_{i}^{2} = P_{i}$,~~$P_{i} P_{j} = 0$,~and~$\sum_{i} P_{i} = I$.

In addition to separating the solvent motion into modes we also
consider spatial localization. To do this, it is convenient to consider first,
the solvent distribution around the electron center of mass.
Because the transition rate from the first excited 
to the ground state is small for the hydrated electron,
electronic transitions can occur
relatively early as well as late in the trajectory.  Late transitions originate
from a nearly equilibrated excited state, while early ones do not correspond to
any equilibrium configuration of the solvent, but do have a memory of the equilibrated
ground state. Thus, the radial electron -- oxygen pair distribution function 
calculated for the solvent configurations 
immediately preceding the electronic transitions 
(Fig.~\ref{fig:radial_distribution})
is a mixture of the distribution functions of the equilibrated
excited and ground states (which are reproduced in Fig.~\ref{fig:radial_distribution}
from~Fig.~7 of~Ref.~\cite{Ben94}) with the prevailing contribution coming from
the former.  Here and further in the paper, unless explicitly stated otherwise, 
the data represent an ensemble average over twenty non-adiabatic transitions.
The distribution function facilitates the definition of solvation shells.  
Based on Fig.~\ref{fig:radial_distribution}, the radius of the first solvation 
shell is taken to be 4.0~\AA. The shell contains 6.2 water molecules on average.  
The second solvation shell, cut-off at 7.5~\AA, includes 50.3 molecules on average.
It is worth mentioning that the equilibrated excited state electronic wave function 
has the characteristic shape of a p-orbital.$^{48}$  Thus, in most of the
generated trajectories, transitions occurred out of a roughly cylindrical rather 
than spherically symmetrical 
state. Unfortunately, we were unable to profoundly exploit this fact in our discussion,
because a cylindrical pair distribution is a two dimensional distribution that requires
many more data points to achieve statistics comparable with that of the one dimensional
radial distribution.  In section~\ref{sec3a}, we do present a qualitative picture 
reflecting contributions of different regions of the electron -- oxygen cylindrical 
distribution to the transition rate.


\subsection{Instantaneous non-adiabatic transition rate}\label{sec2a}

We define an instantaneous transition rate constant using the Golden Rule expression 
(see e.g. Ref.~\cite{El-Sayed77})
\begin{eqnarray}
k &=& \frac{2 \pi}{\hbar} \frac{|V|^{2}}{\Delta E}
\label{eq:k}
\end{eqnarray}
where $\Delta E$ is the electronic energy gap.  To define the coupling matrix
element $V$ we expand the total Hamiltonian in time for the current instantaneous
geometric configuration of solvent molecules and identify the coupling with
the first order terms:
\begin{eqnarray}
H(t_0 + \Delta t) &=& H(t_0) + \frac{\partial H}{\partial t} \Delta t + 
O[(\Delta t)^2] \nonumber \\
V &=& \langle \beta | \frac{\partial H}{\partial t} \Delta t | \alpha \rangle 
= \langle \beta | \sum_Q \nabla_{Q} H  \dot{Q} \Delta t | \alpha \rangle
= \sum_Q \langle \beta | \nabla_{Q} H | \alpha \rangle \dot{Q} \Delta t
= \sum_Q V_Q
\label{eq:V}
\end{eqnarray}
where $\Delta t =$ 1~fs is the time step of the algorithm and the summation goes
over the solvent degrees of freedom $Q$ defined by Eqs.~(\ref{eq:vib}). 
$|\alpha \rangle $ and $|\beta \rangle $ 
are the adiabatic wave functions for the initial and the final electronic states 
at the beginning of the step: $\langle\beta|H(t_0)|\alpha\rangle=0$.  
The velocities $\dot{Q}$ used in~Eq.~(\ref{eq:V}) 
are those at the beginning of the step.
Each $V_Q$ reflects the contribution of a translational, rotational, or
vibrational mode $Q$ to the promotion of the non-adiabatic
electronic transition.

The instantaneous rate constant $k$ determines how fast the initial state
starts to decay into the final state and is directly related to the transition
probability: 
\begin{eqnarray}
P_{\beta \alpha}(\Delta t) &=& | \langle \beta | \partial / 
\partial t | \alpha \rangle  \Delta t |^2 = | \langle \beta |\nabla_Q | 
\alpha \rangle \dot{Q} \Delta t |^2  \nonumber \\
&=& | \frac{\langle \beta |\nabla_Q H | \alpha 
\rangle}{\Delta E}  \dot{Q} \Delta t |^2 = | V / \Delta E |^2 = 
(\hbar / 2 \pi \Delta E) k
\end{eqnarray}
Here, the first equality expresses the probability of changing states 
over the short time interval $\Delta t$.
The second one is the result of chain differentiation applied to the
time derivative of the adiabatic wave function that depends on time implicitly
via the dependence on nuclear coordinates $Q$. The third equality follows
from the Hellmann-Feynman theorem (see~e.g.~Ref.~\cite{Epstein81}).  The
last two are the corollaries of Eqs~(\ref{eq:V}) and~(\ref{eq:k}).

The non-adiabatic coupling element of Eq.~(\ref{eq:V}) is in general a complex
number $V = |V| e^{i \theta}$, however, it is the magnitude of the total $V$ 
that determines the instantaneous transition
rate (Eq.~\ref{eq:k}).  Nevertheless, the magnitudes of the individual terms $V_Q$
in the sum~(\ref{eq:V}) do not correctly reflect the contributions from various shells
and modes to the total coupling due to the complex phases.
For the present analysis, we eliminated 
the complex part of the total coupling matrix element $V$
by an appropriate rotation in the complex plane, thus making it real. 
We use the real parts of the individual terms in
the sum~(\ref{eq:V}) to analyze the roles of different shells and modes; that is,
\begin{eqnarray}
|V| &=& V e^{i \theta} = \sum_Q V_Q e^{i \theta} = \sum_Q Re \{ V_Q e^{i \theta} \}
\label{eq:reV}
\end{eqnarray}


\subsection{The work done on water molecules by the electron during a
non-adiabatic transition}\label{sec2b}

When the electron changes its state, hopping from the first
excited state to the ground state, it releases energy which has to be 
accommodated by the solvent. An average energy gap between the states at
the time of the transition is about 0.55~eV.$^{48}$
Because the force is determined
self-consistently with the evolution of the electronic wave function, the 
electronic energy change $\Delta E$ is equal to the work done
by this force to accelerate water molecules 
\begin{eqnarray}
\Delta E &=& \sum_Q F_{Q}~\dot{Q}~\Delta t
\label{eq:W}
\end{eqnarray}
with the force given by$^{54}$
\begin{eqnarray}
F_{Q} &=& - Re \{ \langle \beta U^+(t,t_1) | \nabla_Q H | U(t,t_0) \alpha \rangle / 
\langle \beta | U(t_1,t_0) | \alpha \rangle \}
\label{eq:F}
\end{eqnarray}
where $t_0$ and $t_1$ indicate the beginning and the end of a time step
and $U$ is the time development operator.

Another route to analysis of the energy released is to take the molecular 
dynamics step twice --
once allowing for the non-adiabatic transition with coordinates $Q_{na}$
and the second time restricting the dynamics to the original adiabatic electronic 
state with coordinates $Q_a$.  The difference in the kinetic energies of solvent
molecules for the non-adiabatic ($K_{na}$) and adiabatic ($K_a$) steps is approximately 
equal to the work done by the electron during the non-adiabatic transition,
\begin{eqnarray}
\Delta E &\approx& K_{na} - K_{a} = \sum_Q \left( \frac{M_Q \dot{Q}_{na}^{\;2}}{2}
- \frac{M_Q \dot{Q}_{a}^{\;2}}{2} \right)
\label{eq:W1}
\end{eqnarray}
Each term on the right hand side of Eq.~(\ref{eq:W}) or Eq.~(\ref{eq:W1})
gives the amount of energy accepted by the particular solvent mode. 

The time evolution of the energy released by the electron in the transition 
and during the subsequent adiabatic relaxation of the electronic ground state 
can be monitored by computing the kinetic energy stored in solvent modes
along the non-adiabatic trajectory after the transition.  This information
will expose the behavior of different modes during equilibration of the solvent.

\vspace{0.2in}

The equations for the coupling matrix element (Eq.~\ref{eq:V}) and the
electronic energy change (Eqs.~\ref{eq:W} and~\ref{eq:F}) have a very similar
structure. In fact, in the limit of $\Delta t \rightarrow 0$, when the Pechukas
force (Eq.~\ref{eq:F}) reduces to the impulsive force pointing along
the Hamiltonian gradient$^{63}$, these two
expressions become proportional to each other.  Correspondingly, the instantaneous hop
approximation often employed in surface hopping methods~$^{40-42,56,63}$
cannot reveal any differences in
the relative roles of nuclear degrees of freedom in promotion of the non-adiabatic
transition and acceptance of the energy released.  However, the
use of the force, which is self-consistently determined 
during the finite duration transition, makes the analysis
accessible.  
Even so, the similarity of promoting and accepting actions of solvent modes is
inherent in the semiclassical approximation. 
Gas phase theories of radiationless 
transitions, especially those describing the ``large molecule'' limit,
often restrict the summation to one or a few promoting modes and regard
the rest as accepting modes, neglecting the extent to which promoting modes
can also accept energy.$^{8-11}$
It will be shown that a promotion picture which involves at most a few
physically interesting degrees of freedom does not hold for the hydrated electron.


\section{Results and discussion}\label{sec3}

\subsection{Promotion of the non-adiabatic transition}\label{sec3a}

Table~\ref{tab:couplings} shows the decomposition of the
coupling matrix element 
calculated at the time of the non-adiabatic transition and averaged over twenty
such transitions.$^{48}$ 
To calculate the coupling due to a specific shell or mode we simply 
restrict the summation 
in Eq.~(\ref{eq:reV}) to the molecules or degrees of freedom which belong to that 
shell or mode.  The largest contributions to the total non-adiabatic coupling matrix
element come, in decreasing order, from the asymmetric stretch, rotation around 
the Z axis (see Fig.~\ref{fig:vib}) and bending of water molecules.
The contribution of the six first shell molecules is about 35 percent, while the second
solvation shell containing about fifty molecules accounts for 60 percent 
leaving only 5 percent for the
rest of the solvent. It is interesting to note that the contributions
from the different modes of the first shell molecules are more evenly
distributed than those of the second shell molecules.  For example, we see that
the coupling due to the bending vibration is comparable to the coupling due
to the asymmetric stretch for the first shell molecules, while the
asymmetric stretch of the second shell molecules is about ten times more
important than the other two vibrations.  Once a molecule gets
close to the electron it does not matter very much what kind of
motion the molecule undergoes. It efficiently couples the electronic
states.  Vibrations in general are more important than rotations, while 
translations can be safely neglected in the calculation of the total coupling.

In order to understand the difference in the relative contributions of 
various modes, one needs to consider the solvation structure together with
the properties of the electronic wave function. 
The spatial characteristics of the electron density
and the surrounding solvent have been studied in Ref.~\cite{Ben94}.
In that study it was found that the excited state of the hydrated 
electron is p-like in shape, while the ground state is approximately 
described by the s-like wave function.  The p to s-state transition dipole moment
defines the orientation of the $C_\infty$ symmetry axis of the p-state.
Ichiye and coworkers$^{64}$ studied in detail
the origin of interaction between water and simple negatively charged solutes.
They found that for the first solvation shell about eighty percent of the interaction 
energy is of the ion - water molecule dipole character. This type of interaction
dominates for the rest of the solvent, the higher order contributions being 
entirely negligible.  According to this result the second shell water molecules 
and to a large extent the first shell ones point their dipole moments towards the
hydrated electron.  The excited state of the electron forms
an elongated cavity, so that water molecules situated at the cavity's longer
side are able to interact with a larger fraction of the electron density
and, therefore, one expects, contribute most to the coupling matrix element (see 
Fig.~\ref{fig:cyl_coupling} discussed later in this section).  

Turning to the coupling matrix element of interest it can be expressed in a number
of alternative ways. For the present purposes we focus on that in which the
operator is the gradient of the Hamiltonian with respect to the solvent coordinate.
Further, we focus on the electrostatic coupling of the electron and solvent, which
should dominate other terms. For the idealized state symmetries (s and p) here,
it follows that only contributions to the gradient of the electrostatic potential
that are antisymmetrical with respect to the p orbital nodal place will yield
non-vanishing contributions. For a water molecule in its equilibrium
geometry, one finds that based only on symmetry (see Fig.~\ref{fig:vib})
the derivative of the dipole moment with respect to displacement is parallel
to the dipole moment for the bend and symmetric stretch coordinate, while
it is perpendicular for the asymmetric stretch. For the present model, the
magnitudes of the derivatives are similar.

Given these facts about solvent-ion interaction and solvent and electronic
organization, along with the dipole derivatives, it follows that the asymmetric
stretching motion of the laterally located molecules produce an electric dipole 
moment parallel to the p-orbital axis and, thus, strongly couples the electronic states.
On the other hand, the other two vibrations vary molecular dipole in this direction 
only for the molecules located at the ends of the lobes of the
electronic wave function. For this reason, the role of the symmetric stretch
and bend is less pronounced.  Bend is slightly more important
that the symmetric stretch, since the molecular dipole derivative with respect
to the bending normal mode is somewhat larger.  These trends are more clear
for the second shell, where dipolar solvent orientation is more 
dominant.$^{64}$
The same argument applied to rotations
correctly predicts the small influence of the rotation around the X-axis
(for the definition of the coordinate system see Fig.~\ref{fig:vib}), since
this motion does not change the direction of the molecular dipole.
The greater role of the Z-axis rotation compared to the Y-axis 
can be understood by comparing the corresponding moments of inertia.
Their ratio is 1:3.  The libration around the Z-axis proceeds
with a greater value of the instantaneous
velocity entering the expression for the coupling matrix element 
(Eq.~\ref{eq:V}).

Table~\ref{tab:av} gives average absolute magnitudes of coupling per nuclear
degree of freedom of the first and second shell molecules given by
(cf. Eqs.~\ref{eq:V} and~\ref{eq:reV}) 
\begin{eqnarray}
V^{abs}_N &=& \frac{1}{N} \sum_{Q=1}^N | \langle \beta | \nabla_{Q} H | \alpha \rangle
\dot{Q} \Delta t |
\label{eq:absV}
\end{eqnarray}
where N is the number of degrees of freedom within the corresponding shell and/or mode.
The absolute magnitude of coupling per degree of freedom
shows how strongly a single first or second shell degree of freedom 
would couple adiabatic electronic states if there were no other modes. 
According to these results, the first shell modes
are about two and a half times more effective than the corresponding
degrees of freedom of the second shell.  

The instantaneous transition rates calculated via Eqs.~(\ref{eq:k}) and~(\ref{eq:V})
are given in Table~\ref{tab:rates}.  Although, as expected from Eq.~(\ref{eq:k}),
the data remains qualitatively similar to the data of Tables~\ref{tab:couplings} 
and~\ref{tab:av}, the quantitative difference is striking.  The asymmetric
stretch alone reproduces half of the transition rate, standing dramatically out from
all other motions, which seem far less important.  
This ``discrepancy'' is explained by the presence of the sign to the coupling matrix
elements.
The coupling matrix elements of Table~\ref{tab:couplings} were obtained by averaging over
twenty transitions, and in some cases the contributions to the averages had 
alternating signs.  To calculate the data of Table~\ref{tab:rates}, we averaged 
over the transition rates proportional to the squares of the 
couplings according to Eq.~(\ref{eq:k}).
The asymmetric stretch always effectively coupled the
electronic states, but in one fifth of all the transitions it 
contributed to the
total coupling matrix element of Eq.~(\ref{eq:reV}) destructively. 
Destructive contributions due to, for instance, the rotation around the Z~axis 
were also observed, although four times less frequently then for the asymmetric
stretch. 

The constructive and destructive contributions to
the total coupling is brought about by the distinctly multi-mode origin of promotion of 
the electronic transition.
Numerous quantum mechanical treatments of gas phase vibronic 
relaxation consider a single mode approximation for induced mixing for systems of
many degrees of freedom (see~Ref.~\cite{Freed76,El-Sayed77,Lin80,Fisher84}
and references therein). This is clearly not valid for the case of 
the hydrated electron. 
Tables~(\ref{tab:couplings})-(\ref{tab:rates}) show that
the transition is not localized in any respect.
In a single mode promotion, there is one nuclear degree of freedom
responsible for the mixing of electronic states, and the coupling matrix
element reduces to a product of two scalars --- the nuclear velocity and the
corresponding derivative of the electronic Hamiltonian.  
The greater the gradient and the velocity, the faster the transition.  
Multi-mode systems can exhibit a nonintuitive behavior:
the coupling can entirely vanish even if the electronic Hamiltonian gradients and 
nuclear velocities are large.  The coupling vanishes when nuclear and electronic 
dynamics are uncorrelated, that is, when nuclear velocities are distributed at random 
with respect to the directions of the corresponding gradients of the electronic 
Hamiltonian.  This follows from the fact that the coupling matrix element of 
Eq.~(\ref{eq:V}) can be regarded as a scalar product of two vectors defined in 
the configuration space of nuclear degrees of freedom
\begin{eqnarray}
V = \langle \nabla H | \dot{R} \rangle  \Delta t;~~
\langle \nabla H | = \sum_Q \langle \beta | \nabla_Q H | \alpha \rangle 
\langle \hat{Q} |,~
| \dot{R} \rangle = \sum_Q~\dot{Q}~| \hat{Q} \rangle
\label{eq:V1}
\end{eqnarray}
where $\hat{Q}$ is the unit vector along the $Q${\em th} nuclear coordinate.
Thus, it is evident that the angle between the gradient of the electronic 
Hamiltonian $| \nabla H \rangle$
and the nuclear velocity $| \dot{R} \rangle$ vectors has a crucial influence
on the magnitude of the coupling.  
Table~\ref{tab:angle} gives the deviation of the
angle from $90^{\circ}$ for different nuclear modes and solvation shells. 
The angles reported are obtained as $\cos^{-1} \{ \overline{V} / \overline{\dot{R}}~
\overline{\nabla H} \}$ with bars above the letters indicating averaging over
twenty transitions.  The significance is large;
if the first solvation shell performed a collective motion acting as a single
mode its contribution to the transition rate would be fifty times greater
($\cos [90.0-1.4] \approx 1/50$). This effect is
an order of magnitude stronger for the second shell.
Among the various modes, the asymmetric stretch is most correlated with the 
electron dynamics explaining why it accounts for a large fraction of the instantaneous
transition rate (Table~\ref{tab:rates}).

The correlation between nuclear velocities and the gradient of the electronic 
Hamiltonian
appears due to the coupling between the electronic and nuclear dynamics, which is
mainly determined by the force (Eq.~\ref{eq:F}) exerted by the electron on solvent 
molecules.  
Table~\ref{tab:forces} presents average magnitudes of the force per 
nuclear degree of freedom for various shells and modes at the transition step,
defined by
\begin{eqnarray}
F^{abs}_N &=& \frac{1}{N} \sum_{Q=1}^N | F_Q |
\label{eq:absF}
\end{eqnarray}
where $F_Q$ is given by Eq.~(\ref{eq:F}) and $N$ is the number of modes in
a particular average.  There is an obvious correlation
between the data of Tables~\ref{tab:angle} and~\ref{tab:forces} supporting
our previous statement.

Fig.~\ref{fig:radial_coupling} shows the radial electron -- oxygen
pair coupling distribution $v(R)$ together with the pair distribution of 
coupling magnitude $v^{abs}(R)$, where $R$ is the scalar distance between
the oxygen nucleus and the electronic center of mass.  The integral of 
$v(R)$ weighed by the radial pair distribution function $g(R)$ 
(Fig.~\ref{fig:radial_distribution}) gives the magnitudes of the total coupling 
matrix element $|V|$ of Eq.~(\ref{eq:reV}):
\begin{eqnarray}
|V| &=& \int_{0}^{\infty} dR~4 \pi R^{2}~g(R)~v(R)~\rho \nonumber \\
\sum_Q~|V_Q| &=& \int_{0}^{\infty} dR~4 \pi R^{2}~g(R)~v(R)^{abs}~\rho 
\label{eq:v}
\end{eqnarray}
with $\rho$ being the solvent density.
The corresponding integral of $v^{abs}(R)$ results in 
the sum of absolute values of $V_Q$ defined in Eq.~(\ref{eq:V}).
The absolute magnitude of the coupling is at a maximum in the first solvation
shell and smoothly decays with the distance $R$.  The coupling almost vanishes for 
the molecules which are situated on the border between the first and the 
second solvation shells.  This can be explained as follows.
The molecules within the first and the second shell differ in their
orientation toward the electron, and, hence, the border region exhibits
a less ordered structure.  Although a single molecule
of the border region makes a significant contribution to the
total coupling matrix element, on the average the net contribution
becomes negligible due to the variations in molecular orientation.

The functions $v(R)$ and $v^{abs}(R)$ reflect the coupling due to a single
water molecule at the distance $R$ from the electron center of mass.  
While these functions decrease with $R$, the number of molecules increases.
To visualize the role of all molecules located at the distance $R$ 
from the electron we plot the distribution of the cumulative coupling
(Fig.~\ref{fig:radial_coupling_t}) defined as
\begin{eqnarray}
v^{cum}(R) &=& 4 \pi R^{2}~v(R)~g(R)~\rho   \nonumber \\
|V| &=& \int_{0}^{\infty} dR~v^{cum}(R)  
\label{eq:vcum}
\end{eqnarray}
Although a single molecule of the first solvation shell couples the electronic 
states much more strongly than a single molecule of the second shell 
(Table~\ref{tab:av}, Fig.~\ref{fig:radial_coupling}), 
the cumulative contribution of the second
shell is twice that of the first shell (Table~\ref{tab:couplings}, 
Fig.~\ref{fig:radial_coupling_t}). At large distances the radial electron -- oxygen
pair distribution of the coupling per molecule 
(Fig.~\ref{fig:radial_coupling}) decays faster then $1/R^2$ so that the role of
the molecules beyond the second solvation shell is insignificant.

Since the excited state electronic wavefunction is p-like in shape, it is more
cylindrically than spherically symmetric. To better illustrate how
different spatial regions of the solvent promote the electronic transition
we display a contour plot of the cylindrical pair distribution function
of coupling $v^{cyl}(R,|Z|)$ (Fig.~\ref{fig:cyl_coupling}). The Z-axis 
of the distribution is defined by the direction of the first excited to the
ground state transition dipole moment and parallels the $C_\infty$ approximate symmetry
axis of the p-like excited state.  Because the two dimensional cylindrical distribution 
has inherently poorer statistics than the one dimensional radial
distribution, in Fig.~\ref{fig:cyl_coupling} we coarse-grain the coordinate axes 
by 0.1\AA~and then compute running averages over sets of 20 adjacent points. 
We also average over the sign of the Z coordinate.
The cylindrical distribution of coupling $v^{cyl}(R,|Z|)$ is defined in such a way,
that its integral weighed by the corresponding cylindrical electron -- oxygen pair
distribution $g^{cyl}(R,Z)$ gives the magnitude of the total coupling matrix element 
of Eq.~(\ref{eq:reV}):
\begin{eqnarray}
|V| &=& \int_{0}^{\infty} 2 \pi R~dR \int_{-\infty}^{\infty} 
dZ~g^{cyl}(R,Z)~\frac{v^{cyl}(R,|Z|)}{2}~\rho
\label{eq:vcyl}
\end{eqnarray}
We recall that the electron density for p-state is elongated. The distribution
of coupling evident in Fig.~\ref{fig:cyl_coupling} is consistent with the expectation
that solvent located centrally with respect to this elongated electron density
is more strongly coupled than that near only to one end ($Z \sim \pm 5$\AA).
Further, it is reasonable to expect the effect of the changing solvent
induced electric field is of particular importance due to the presence
of an electronic nodal region.

Finishing our discussion of the promotion of the non-adiabatic electronic
transition we present Figure~\ref{fig:long_promotion}, which shows the temporal
evolution of the instantaneous rate (Eq.~\ref{eq:k}) of the downward 
transition over one hundred femtoseconds before the hop as well as the instantaneous
rate of the reverse process for the forty femtoseconds after it. 
(Note the logarithmic scale of the y-axis.)
The figure is based on the data from a single trajectory. 
We observe two time scales in the fluctuation of the transition
rate with periods of about ten and fifty femtoseconds, which correspond 
to the vibrational and librational motions of the solvent. Vibrations cause fluctuations
of much larger amplitude. During each vibrational period, 
the rate is high for two or three femtoseconds, when the time derivative
of the vibrational coordinate $\dot{Q}$ is large (cf. Eq.~\ref{eq:V}).
It is interesting that the probability to make the downward 
transition does not show any trends during the period prior
to the transition point, with a fluctuation occurring at roughly 50~fs
leading to a transition probability nearly as large as that acting at 
the observed transition point.
However, once the transition occurs the probability for an upward transition
rapidly decays; this decay of the instantaneous transition 
rate (Eq.~\ref{eq:k}) of the reverse process is due to the rapid separation of the 
electronic energy levels ($\Delta E$) brought about by the fast inertial 
response of the solvent immediately following the transition.$^{48}$ 


\subsection{Energy transfer during and after the non-adiabatic transition}
\label{sec3b}

During the one femtosecond step of a non-adiabatic transition the electronic energy
abruptly changes by about $0.5$~eV.  This excess energy is transferred to
the solvent leading to  local heating.  
Tables~\ref{tab:work} and~\ref{tab:energy_difference} show the energies
accepted by different shells and modes calculated by the formulas (\ref{eq:W}) 
and (\ref{eq:W1}) of section~\ref{sec2b}, respectively.
Both tables show similar data, with the minor discrepancies being 
mainly due to the finite difference scheme used in the molecular
dynamics algorithm to calculate solvent molecules velocities. 
About forty percent of the total energy disposed by
the electron is transferred to the first solvation shell.
The second shell accepts another fifty percent.
As in the case with the promotion discussed in the previous subsection,
energy release during the non-adiabatic transition of the hydrated electron is not 
localized but involves both the first and the second shells.  
Nevertheless, the relative impact of the electronic
transition on the first shell molecules is substantial, leading to an initial
local heating of $40$~K. Over the next several femtoseconds
the electron experiences a rapid adiabatic relaxation, continuing to release energy 
and the local heating reaches almost $100$~K (see Fig.~\ref{fig:long_acceptance} 
discussed later in this section).  We emphasize that this local heating
does not have such dramatic consequences on reorientational and diffusive 
motions in the solvent as a uniform heating would have.  The overall temperature
rise is only several Kelvin and the relaxation dynamics proceeds on a usual 
time-scale.$^{34}$
For instance, it takes several hundred femtoseconds until the first and
the second shells equilibrate. (See Ref.~\cite{Maroncelli88} for a discussion
of the influence of local heating on solvation dynamics.)
Within the first solvation shell, it is the asymmetric stretch
which gets most agitated during the transition. Both
the asymmetric stretch and the rotation around the Z~axis 
(see Fig.~\ref{fig:vib}) of an average solvent molecule each equally 
account for one third of the non-adiabatic electronic energy change.  
The rotation around the Z~axis is most involved in part because it corresponds
to the smallest moment of inertia.  It is easily shown, 
that if two rotors hold equal amounts of energy $E_1 = E_2$ (equipartition 
ansatz) and if the torques acting on the rotors are the same $\dot{M}_1 
= \dot{M}_2$, the energy transfer is faster for the rotor with the smaller
moment of inertia:
\begin{eqnarray}
\frac{dE_1/dt}{dE_2/dt}&=&\frac{\frac{d}{dt}(M_1^{\,2}/2I_1)}
{\frac{d}{dt}(M_2^{\,2}/2I_2)} = \frac{(M_1/I_1)\dot{M_1}}
{(M_2/I_2)\dot{M_2}} = \sqrt{\frac{I_2}{I_1}}
\nonumber
\end{eqnarray}
Here $I$ and $M$ denote the moment of inertia and the angular momentum
correspondingly.

The next two figures show the radial electron -- oxygen pair distributions
of energy accepted during the non-adiabatic transition by a single
molecule $\varepsilon$ (Fig.~\ref{fig:radial_energy})
and by all molecules $\varepsilon^{cum}$ (Fig.~\ref{fig:radial_energy_t}) at
the distance $R$ from the electron.  These distribution functions are
defined analogously to the corresponding radial distributions of coupling
(Eqs~(\ref{eq:v}) and~(\ref{eq:vcum}), where $v$, $v^{cum}$ and $V$
correspond to $\varepsilon$, $\varepsilon^{cum}$ and the total 
accepted energy $\Delta E$).
In general, the closer a molecule is to the electron, the more energy it accepts.
The decay of the distribution function in Fig.~\ref{fig:radial_energy}
with distance is sufficiently gradual, that the cumulative effect of molecules 
as far as $6$~\AA~from
the electron center of mass is still significant (Fig.~\ref{fig:radial_energy_t}).  
Comparing the radial distribution of energy accepted during the transition 
(Figs~\ref{fig:radial_energy} and \ref{fig:radial_energy_t}) 
with the distribution of coupling between the electronic states 
(Figs~\ref{fig:radial_coupling} and \ref{fig:radial_coupling_t}) 
we note that the former does not have the minimum around $4$~\AA~(in the 
intershell region) which was quite noticeable in the latter.  
During the non-adiabatic transition the electron strongly acts on all surrounding water
molecules independent of their location. The tenfold difference
in the magnitudes of the coupling matrix element ($\simeq0.05$~eV) and the transferred 
energy ($\simeq0.5$~eV) explains this observation.

After the electron makes the transition to the ground state, it continues to
release energy adiabatically due to solvation on the newly formed ground state.  
Calculations predict two time-scales for this equilibration process.$^{48}$ 
The first part of it is due to the inertial solvent response. It lasts for
about $20$~fs, during which the electron loses most of the $2$~eV of the total 
lowering of the ground adiabatic state. 
The second longer part extends for about $250$~fs. 
Temporal variation of the energy contained within the first and 
the second solvation shells is given in Figure~\ref{fig:long_acceptance}, 
where we plot local temperature as a function of time after the transition.  
The first solvation shell undergoes a strong heating for $10$~fs.
The second shell becomes hotter to a significantly 
lesser extent. Further, its response is mostly secondary, responding to
the first shell cooling, transferring energy to the second
shell and to the rest of the solvent.
It is the delay in response of the second shell rather than the
fast heating of the first one that matches the $20$~fs inertial part of the
relaxation. The first shell water molecules vigorously accelerate during the first 
$10$~fs after the transition and continue to move very fast for another $10$~fs 
adjusting themselves on the new adiabatic potential surface. More global
changes in solvent structure, including second shell molecules,
are responsible for the slower component of the hydrated electron ground state 
relaxation.

Various degrees of freedom within the first solvation shell do not respond equally
to the adiabatic electronic state relaxation.
For several dozen femtoseconds after the non-adiabatic
transition, rotations hold a much larger fraction of the energy than vibrations
and translations. 
Figure~\ref{fig:long_acceptance1} demonstrates this.
An understanding of this comes from
the difference in the frequencies of vibrational and librational motions 
of bulk water.
Water molecules complete their vibrational cycle within ten femtoseconds, 
during half of which molecule velocity vectors point in the direction opposite
to the force due to the electron. Because the electron continues rapid adiabatic
relaxation longer than half of the vibrational period, the accepted energy 
averaged over the period is small.
On the contrary, the librational period is about fifty femtoseconds. 
Twenty femtosecond intensive energy discharge covers less than half of this.  
The force -- velocity dot product does not oscillate and the contribution
of librations to the total work of Eq.~(\ref{eq:W}) is large. 
Thus, the librations are primarily responsible for effective energy transfer.


\section{Conclusions}\label{sec4}
In summary, our theoretical study on the involvement of particular solvent
degrees of freedom in non-radiative relaxation of the hydrated electron
has shown that the electron relaxes due to a spatially delocalized
multi-mode coupling between the adiabatic states. 
The largest contribution to the coupling comes from
the asymmetric stretch of water molecules. Bending vibration and rotation around 
symmetry axis are also important degrees of freedom.
The transition rate is not determined by motions solely within the first 
solvation shell, the second shell contribution being twice as large.  
The temporal behavior of the transition probability exhibits 
two time scales corresponding to vibrational and librational bands of the bulk 
water spectrum. 

The evolution of energy released during the transition
proceeds in two steps. The first one takes about $10$~fs 
and is characterized by a vigorous heating of the first solvation shell.
The second one involves global reorganization of solvent structure and
lasts several hundred femtoseconds. Energy
transferred to the first shell is accepted primarily by solvent librations.
The reason for this is a good match between the libration period and the duration
of adiabatic electronic energy loss following the transition. Therefore, in short, 
vibrations stimulate the hydrated electron change of state, 
while first shell librations are
responsible for immediate acceptance of released energy.
However, in both cases, it is the proton motion that is dominant.

Studies of relaxation for more complex solutes will be of interest
in determining the generality of these results, and the division between 
intramolecular and solvent modes.


\section*{Acknowledgments}\label{sec5}
This work has been supported by a grant from the National Science Foundation.
We also acknowledge computational support of the UT-Austin High Performance
Computing Facility. We thank Dr.~Benjamin J. Schwartz and Dr.~Eric R. Bittner
for useful comments and discussion.



\newpage 

\begin{table}[h]
\caption{Components of coupling matrix element (meV) between the first excited 
and ground states (Eqs.~\ref{eq:V}~and~\ref{eq:reV}).}
\label{tab:couplings}
\vspace{.1in}
\begin{tabular}[t]{l|*{8}c} \hline
due to  & trans & rot[x] & rot[y] & rot[z] & v [a] & v [s] & v [bd]  & all modes \\
  \hline \\
  1st shell     & 0 & 1 & 3 &  4 &  6 & 1 & 4 & 19 \\
  2nd shell     & 0 & 1 & 2 & 13 & 14 & 0 & 2 & 32 \\
  all molecules & 0 & 3 & 6 & 14 & 20 & 2 & 9 & 54 \\  
  \hline
\end{tabular}
\vspace{1in}
\end{table}

\begin{table}[p]
\caption{Average absolute magnitudes of coupling (meV) per nuclear 
degree of freedom (Eq.~\ref{eq:absV}) between the first excited and ground states.} 
\label{tab:av}
\vspace{.1in}
\begin{tabular}[t]{l|*{8}c} \hline
due to  & trans & rot[x] & rot[y] & rot[z] & v [a] & v [s] & v [bd]  & all modes \\
  \hline \\
  1st shell     & 0.1 & 0.7 & 2.2 & 3.5 & 3.3 & 1.7 & 1.1 & 1.4 \\
  2nd shell     & 0.1 & 0.3 & 0.7 & 1.5 & 1.4 & 0.6 & 0.4 & 0.6 \\
  \hline
\end{tabular}
\vspace{1in}
\end{table}

\begin{table}[p]
\caption{Instantaneous rates (ps$^{-1}$) of the first excited to 
ground state transition at the step of the non-adiabatic transition
calculated from Eqs.~(\ref{eq:k}),~(\ref{eq:V}) and averaged over 20 transitions.}
\label{tab:rates}
\vspace{.1in}
\begin{tabular}[t]{l|*{8}c} \hline
due to  & trans & rot[x] & rot[y] & rot[z] & v [a] & v [s] & v [bd]  & all modes \\
  \hline \\
  1st shell     & 0.00 & 0.00 & 0.01 & 0.00 & 0.02 & 0.01 & 0.01 & 1.77 \\
  2nd shell     & 0.00 & 0.00 & 0.07 & 0.02 & 0.56 & 0.05 & 0.05 & 5.93 \\
  all molecules & 0.00 & 0.08 & 0.08 & 0.22 & 5.94 & 0.08 & 0.74 & 12.08 \\  
  \hline
\end{tabular}
\vspace{1in}
\end{table}

\begin{table}[p]
\caption{Average deviations of the angle (in degrees) between $\nabla H$ and $\dot{Q}$
of Eq.~(\ref{eq:V1}) from the right angle.  The larger the
deviation the more the nuclear dynamics is correlated with the dynamics of 
the electronic structure.}
\label{tab:angle}
\begin{tabular}[t]{l|*{8}c} \hline
  & trans & rot[x] & rot[y] & rot[z] & v [a] & v [s] & v [bd]  & all modes \\
  \hline \\
  1st shell     & 12.8 & 5.8 & 7.8 & 10.8 & 17.9 & 12.0 & 15.9 & 1.4 \\
  2nd shell     &  1.4 & 0.5 & 1.8 &  2.3 &  8.7 &  2.1 &  3.0 & 0.2 \\
  \hline
\end{tabular}
\vspace{1in}
\end{table}

\begin{table}[p]
\caption{Absolute magnitudes of the quantum (Eq.~\ref{eq:F}) force (eV/\AA) per nuclear
degree of freedom, as defined by Eq.~(\ref{eq:absF}).}
\label{tab:forces}
\begin{tabular}[t]{l|*{8}c} \hline
  & trans & rot[x] & rot[y] & rot[z] & v [a] & v [s] & v [bd]  & all modes \\
  \hline \\
  1st shell     & 0.12 & 0.06 & 0.25 & 0.22 & 0.41 & 0.12 & 0.13 & 0.15 \\
  2nd shell     & 0.03 & 0.02 & 0.07 & 0.08 & 0.12 & 0.03 & 0.03 & 0.05 \\
  \hline
\end{tabular}
\vspace{1in}
\end{table}

\begin{table}[p]
\caption{Work (in meV) due to the quantum force
(Eq.~\protect{\ref{eq:F}}) done by the electron on the solvent 
during the one femtosecond non-adiabatic transition calculated via
Eq.~(\protect{\ref{eq:W}}).
}
\label{tab:work}
\begin{tabular}[t]{l|*{8}c} \hline
work done on  & trans & rot[x] & rot[y] & rot[z] & v [a] & v [s] & v [bd]  & all modes \\
~(T change)   &  \\
  \hline \\
1st shell     &  2   &  9   & 22   &  32   &  76   & 17   & 28   & 177   \\  
2nd shell     &  2   &  6   & 23   &  98   &  78   & 17   & 28   & 252   \\  
all molecules &  4   & 17   & 57   & 146   & 158   & 38   & 65   & 485   \\  
  \hline
\end{tabular}
\vspace{1in}
\end{table}

\begin{table}[p]
\caption{Difference in kinetic energies (in meV) resulting from
the one femtosecond non-adiabatic transition calculated 
via Eq.~(\ref{eq:W1}).
Local heating (in K) of the solvent is given in parentheses.}
\label{tab:energy_difference}
\begin{tabular}[t]{l|*{8}c} \hline
work done on  & trans & rot[x] & rot[y] & rot[z] & v [a] & v [s] & v [bd]  & all modes \\
~(T change)   &  \\
  \hline \\
1st shell     &  3   & 13   & 28   &  42   &  76   & 10   & 29   & 200   \\  
              & (2)&(24)&(53)& (75)&(141)&(18)&(62)& (42)\\  
2nd shell     &  4   &  5   & 27   & 107   &  78   &  7   & 32   & 261   \\  
              & (0)& (1)& (6)& (25)& (18)& (2)& (8)&  (7)\\  
all molecules &  7   & 19   & 73   & 173   & 171   & 19   & 64   & 525   \\  
              & (0)& (1)& (4)& (10)&  (10)& (1)& (4)&  (3)\\  
  \hline
\end{tabular}
\vspace{1in}
\end{table}


\newpage

\begin{figure}[p]
\caption{Vibrational modes of a distorted water molecule defined by 
Eqs.~(\protect{\ref{eq:vib}}).~~~~~~~~~~~~~~~~~~~}
\label{fig:vib}
\end{figure}

\begin{figure}
\caption{Electron -- oxygen radial distribution function
for configurations immediately before the electronic transition together with the
corresponding distributions for the equilibrated excited and ground states of the
hydrated electron.}
\label{fig:radial_distribution}
\end{figure}

\begin{figure}
\caption{Radial electron -- oxygen pair distribution 
of coupling per molecule $v(R)$ (solid line) 
and radial distribution of coupling magnitude $v_{abs}(R)$ (dashed line)
of Eqs.~(\ref{eq:v}).}
\label{fig:radial_coupling}
\end{figure}

\begin{figure}
\caption{Radial electron -- oxygen pair distribution 
of the cumulative coupling ($v^{cum}$) from Eq.~(\ref{eq:vcum}).
The integral of $v^{cum}$ is given by the dashed line.}
\label{fig:radial_coupling_t}
\end{figure}

\begin{figure}
\caption{Contour plot of the cylindrical electron -- oxygen pair distribution 
function of coupling between the P-like excited and ground states
of the hydrated electron $v^{cyl}(R,|Z|)$ from Eq.~(\ref{eq:vcyl}). Darker
regions correspond to stronger coupling.  The Z-axis of the plot
corresponds to an approximate $C_{\infty}$ symmetry axis of the ``dumb-bell'' 
shaped excited state electron density.$^{48}$} 
\label{fig:cyl_coupling}
\end{figure}

\begin{figure}
\caption{Instantaneous rate (Eq.~\ref{eq:k}) of the downward (negative
time) and upward (positive time) transitions for a single representative
trajectory.}
\label{fig:long_promotion}
\end{figure}

\begin{figure}
\caption{Radial electron -- oxygen pair distribution 
of energy accepted during the electronic transition by a {\em single} molecule
at the distance $R$ from the electron.}
\label{fig:radial_energy}
\end{figure}

\begin{figure}
\caption{Radial electron -- oxygen pair distribution ($\varepsilon^{cum}$)
of energy accepted during the electronic transition by {\em all} molecules
at the distance $R$ from the electron (solid line). The integral of
$\varepsilon^{cum}$ is given by the dashed line.}
\label{fig:radial_energy_t}
\end{figure}

\begin{figure}
\caption{Energy evolution within the solvent after the non-adiabatic 
electronic transition.
Solid line indicates temperature of the whole solvent, dashed line
gives temperature of the first solvation shell, and dotted line gives 
temperature of the second shell.}
\label{fig:long_acceptance}
\end{figure}

\begin{figure}
\caption{Energy distribution within the first solvation shell as a function
of time since the non-adiabatic transition.  
Solid line gives translational temperature, 
dashed line indicates vibrational temperature, 
and dotted line gives rotational temperature.}
\label{fig:long_acceptance1}
\end{figure}

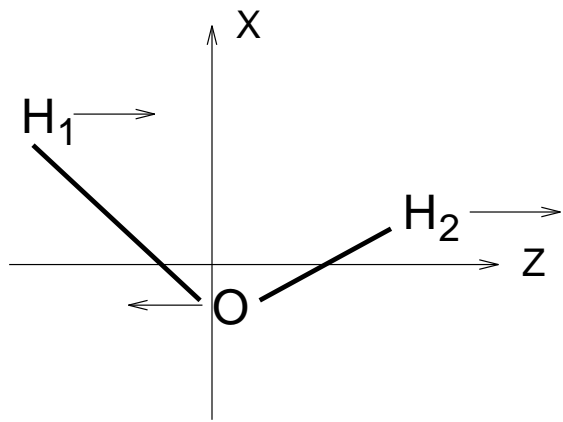
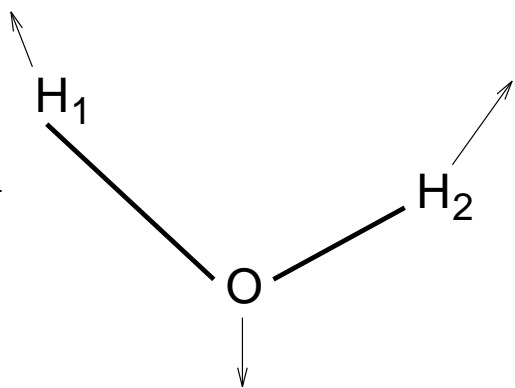
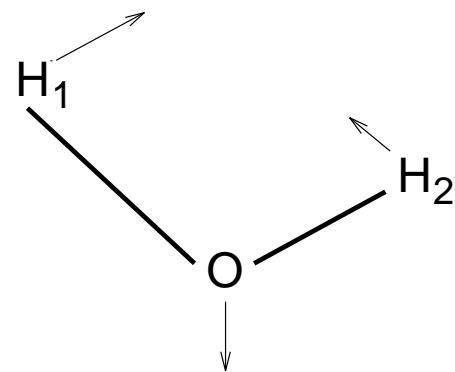
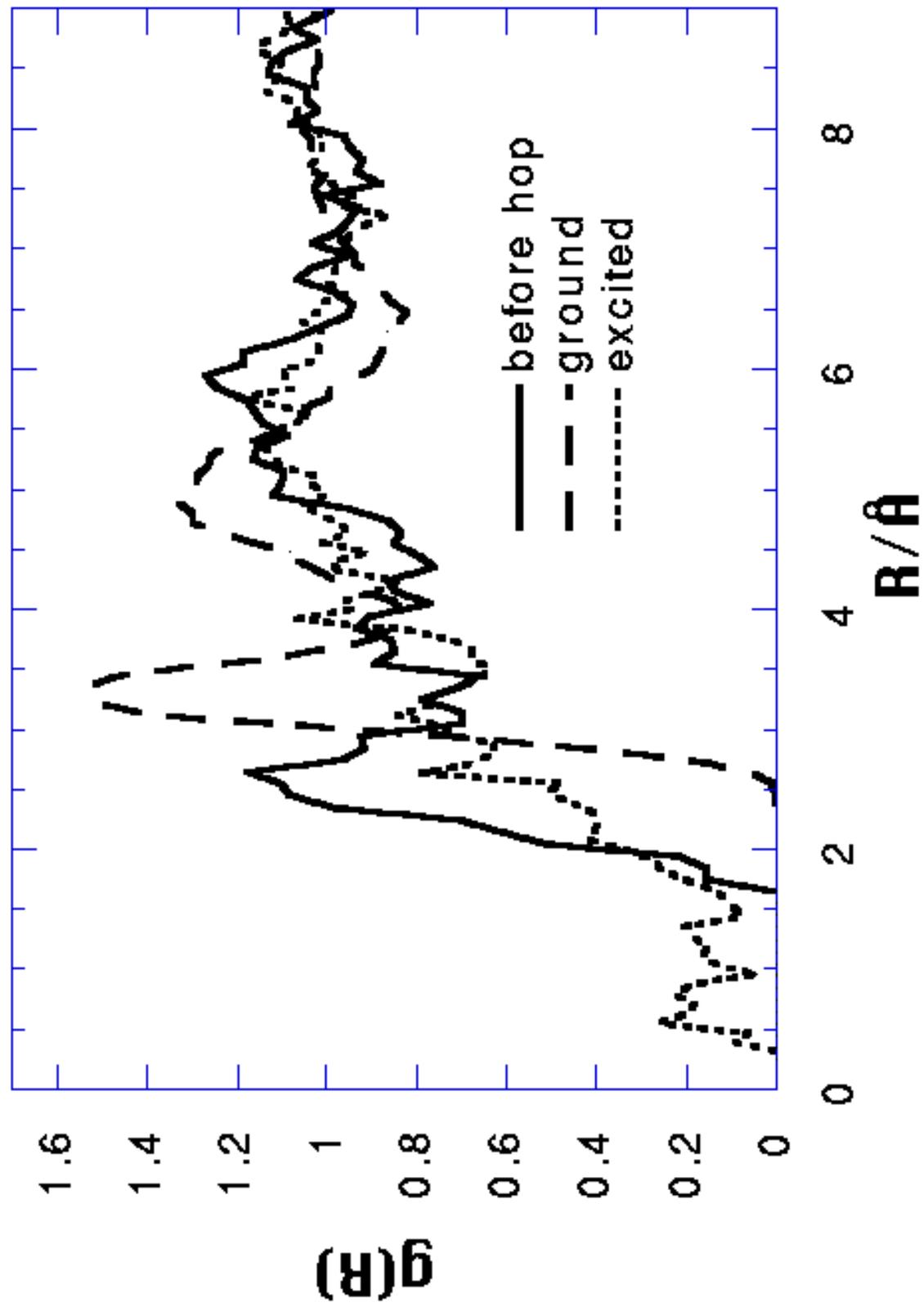
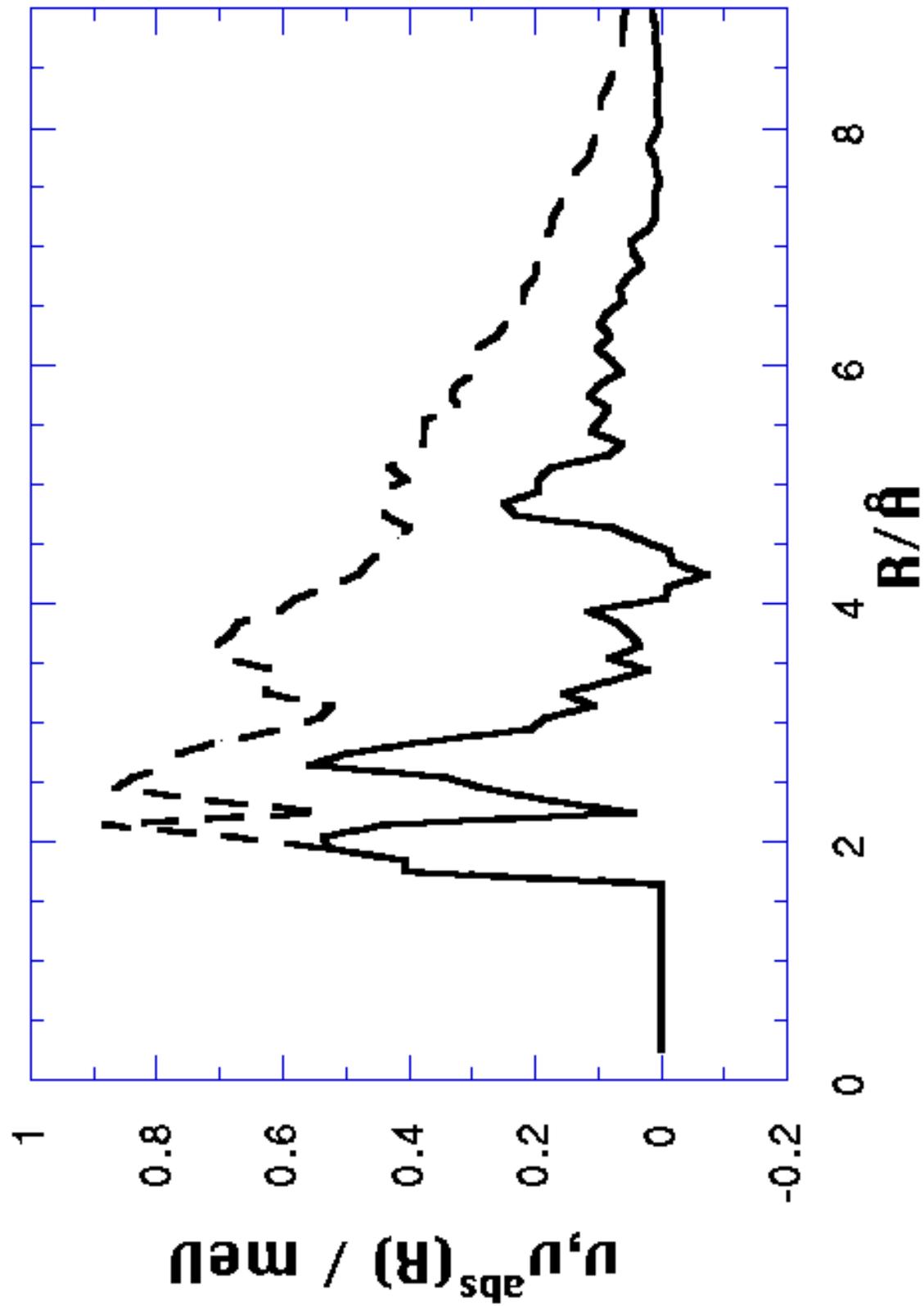
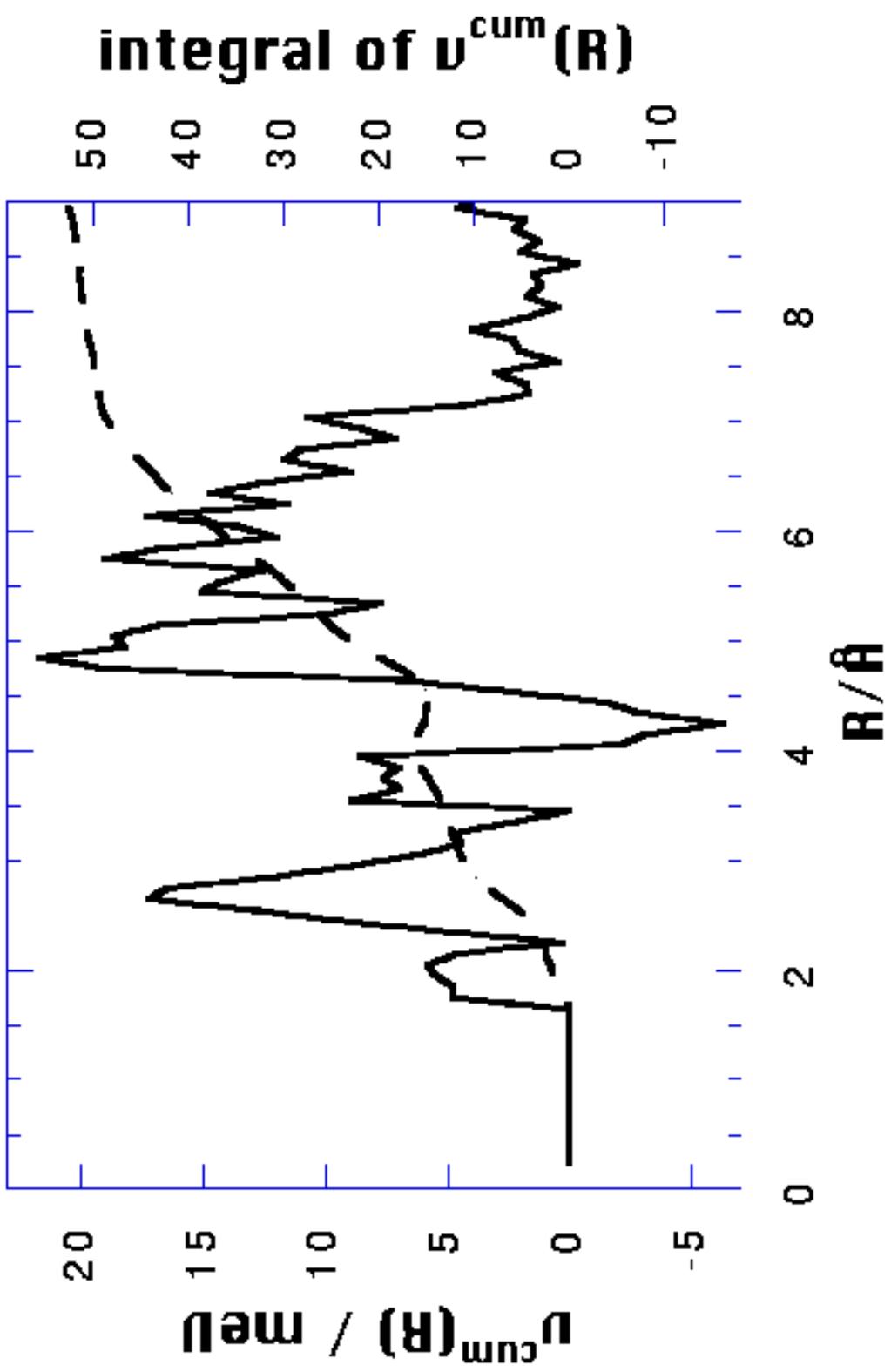
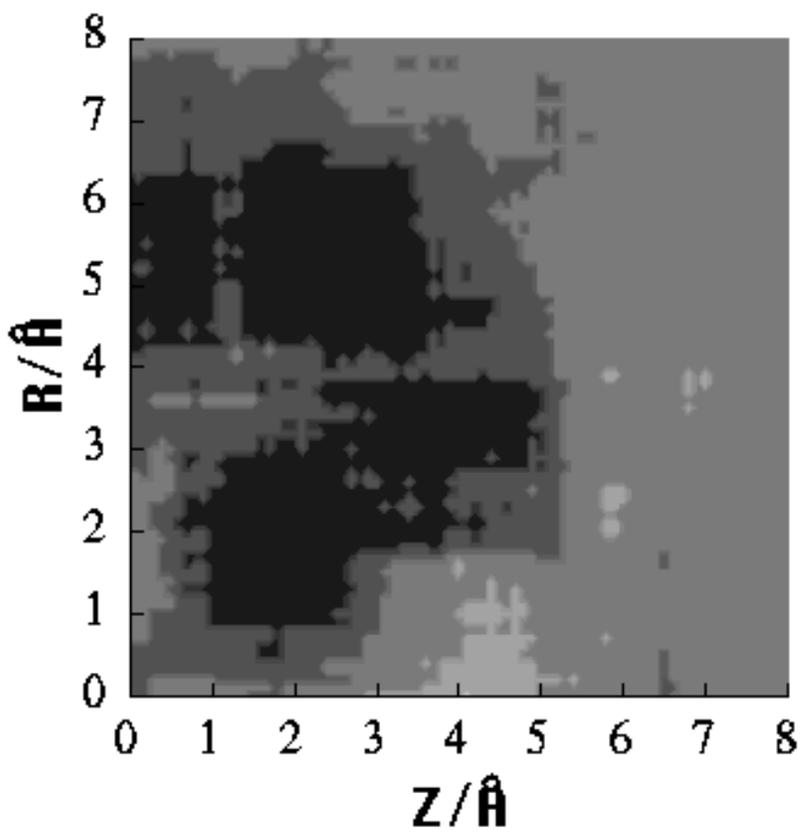
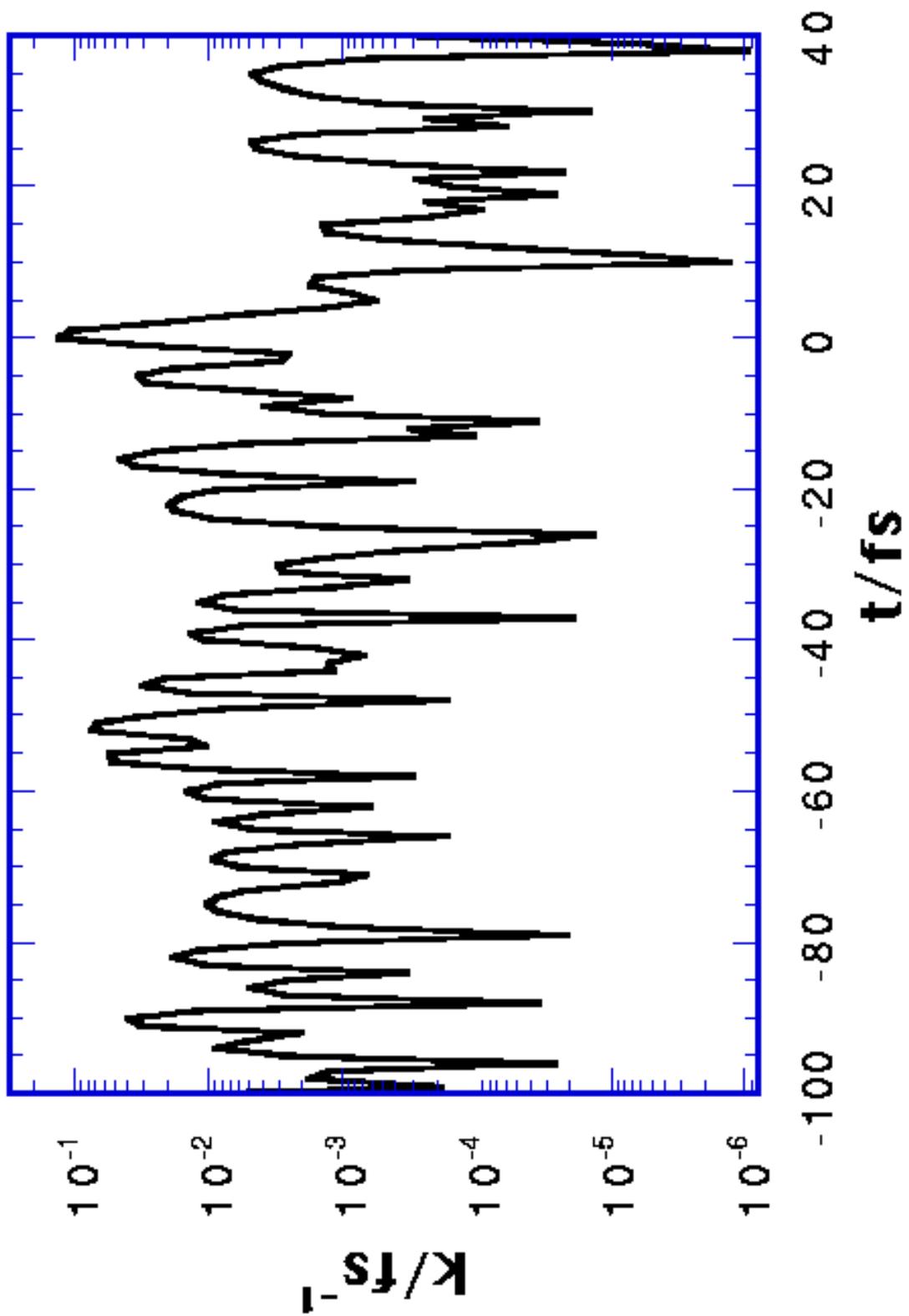
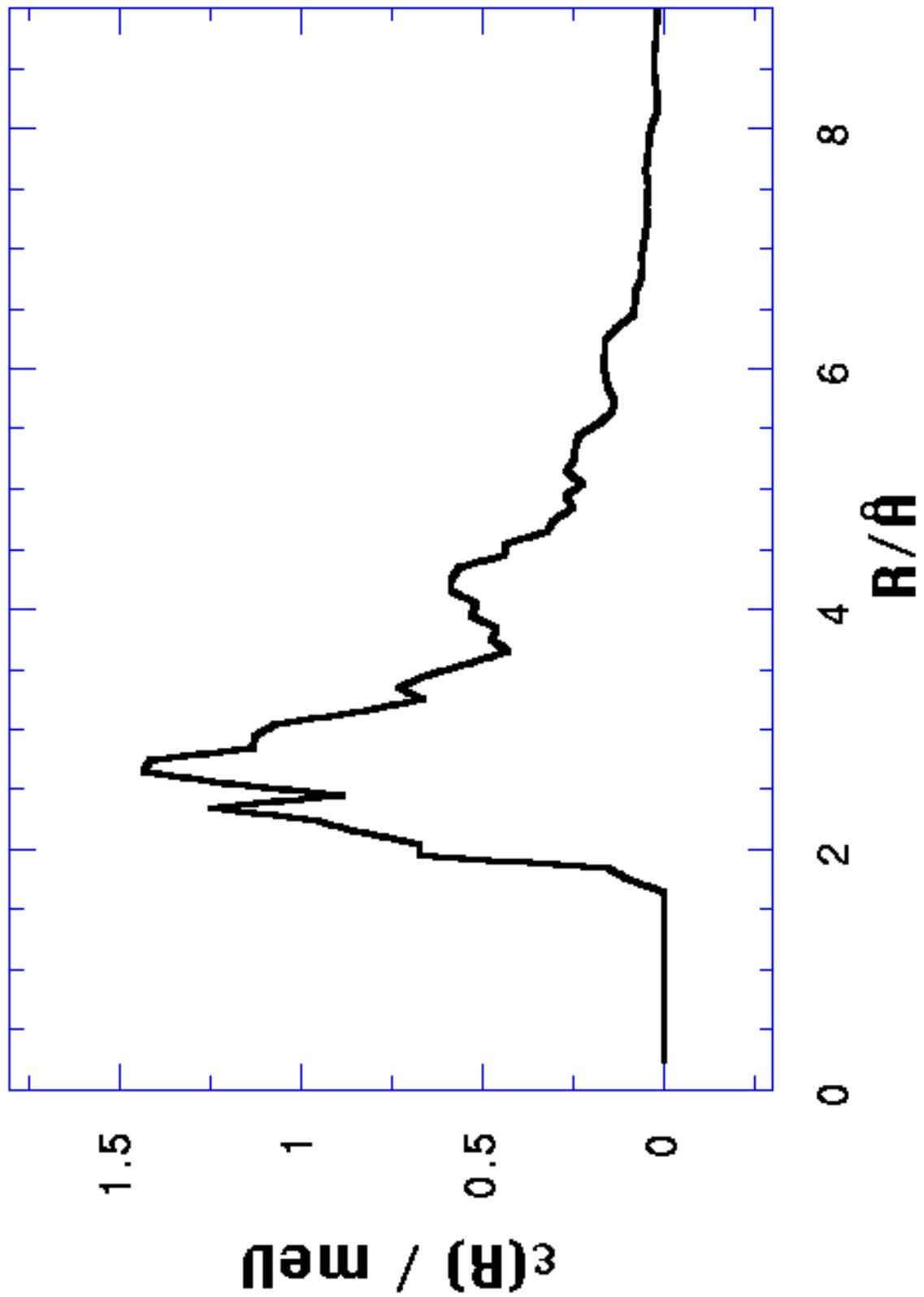
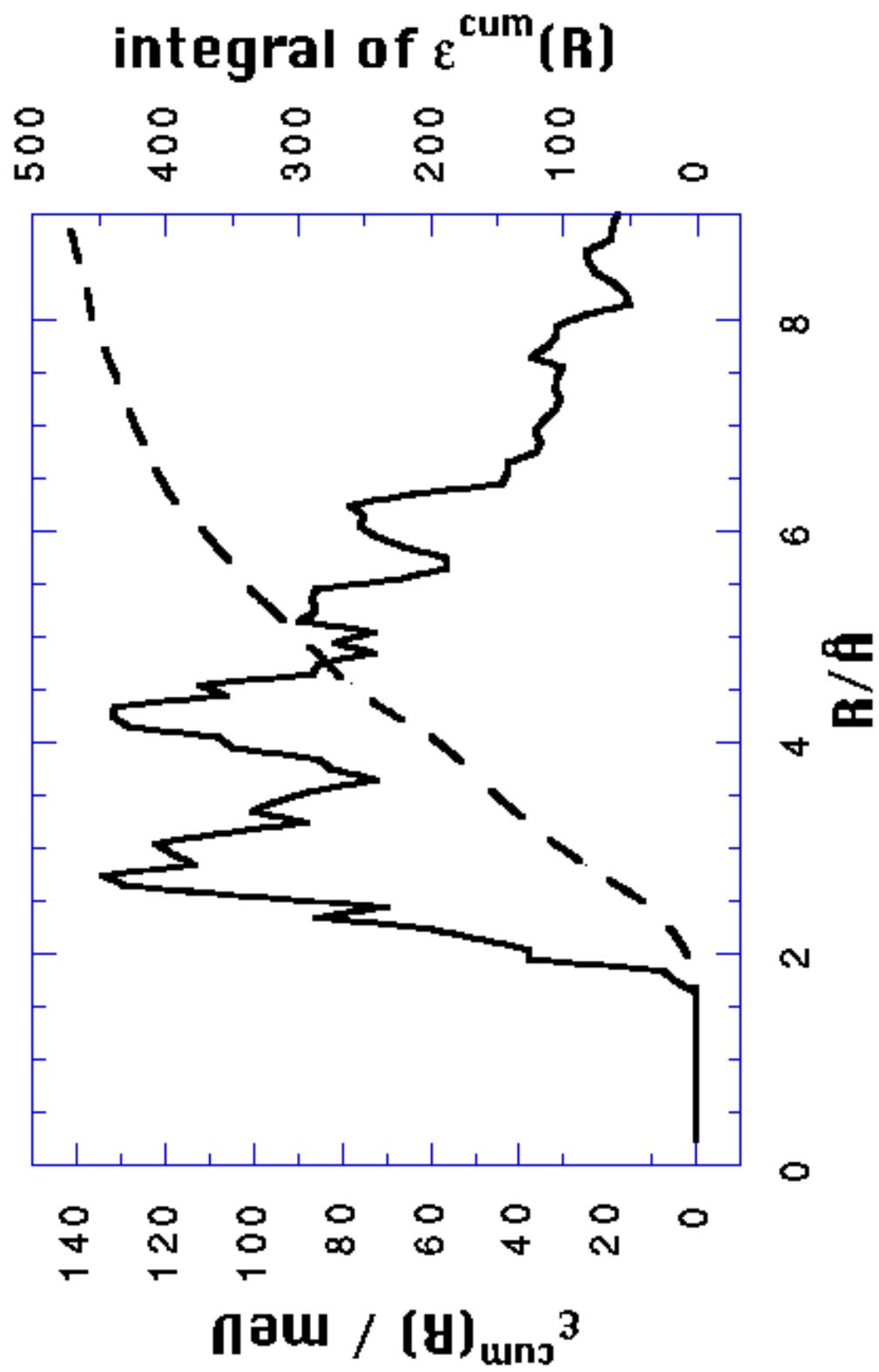
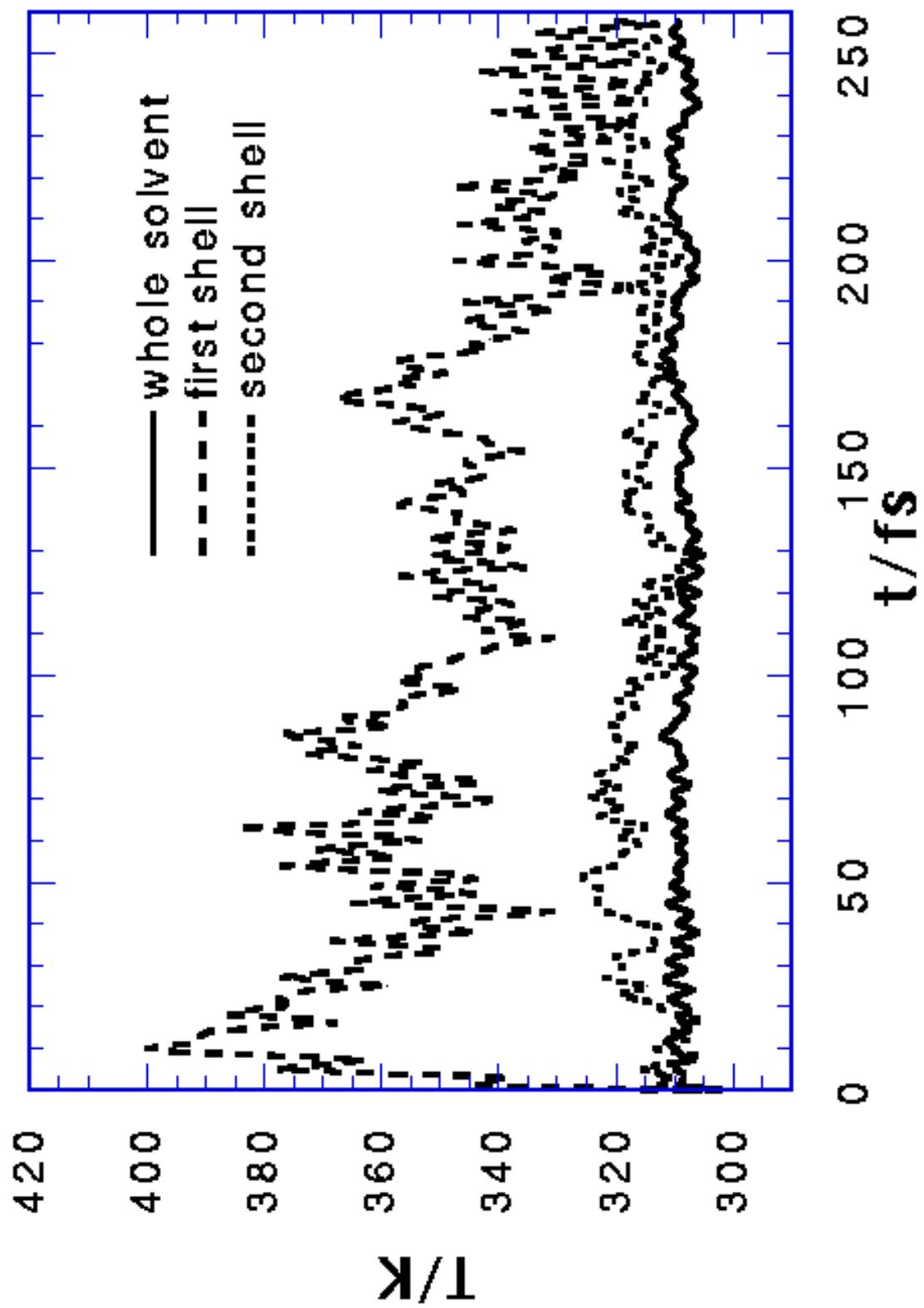
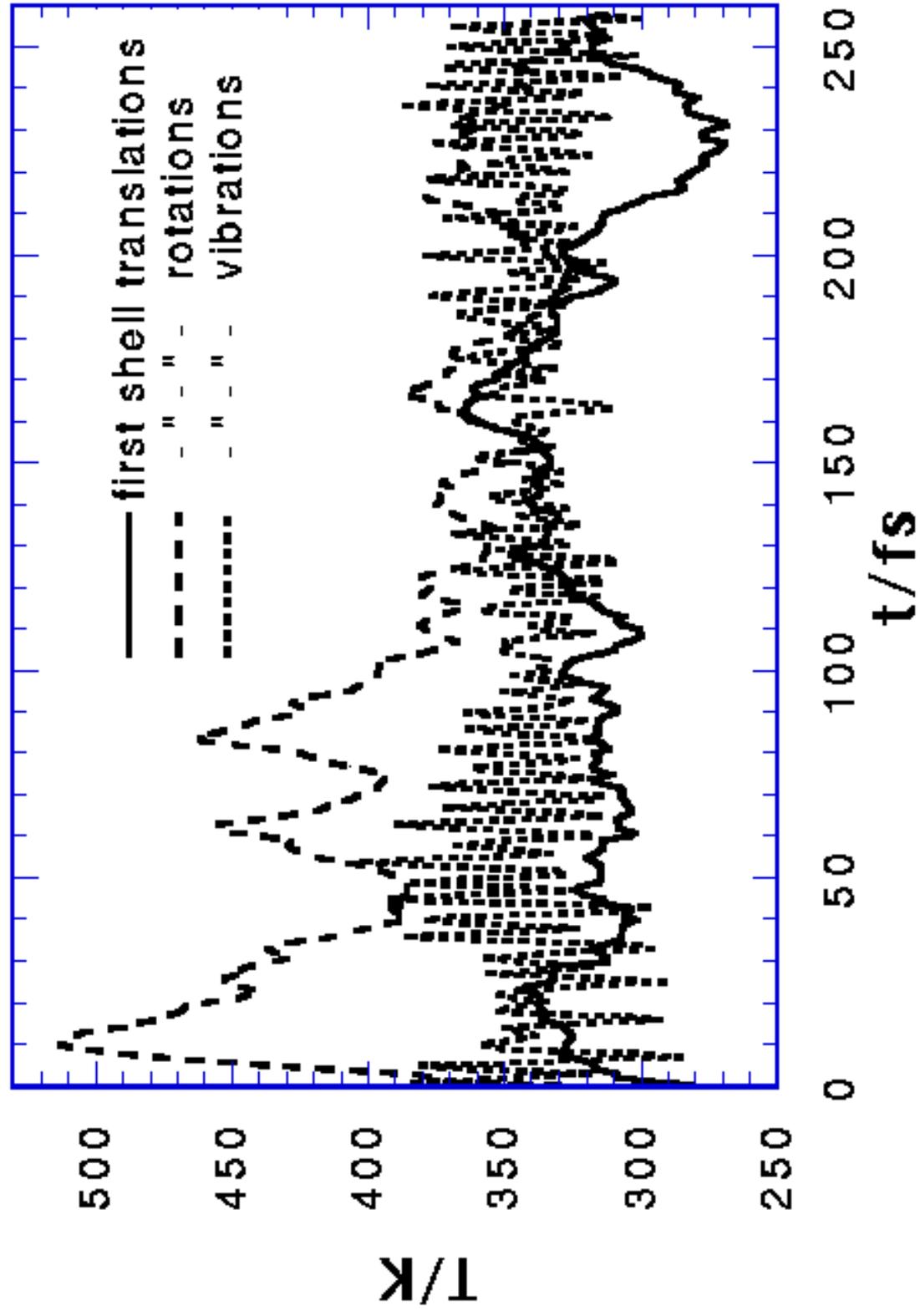
\end{document}